\documentstyle{article}
\advance\textheight by \topskip
\newcommand{\be}{\begin{equation}}
\newcommand{\en}{\end{equation}}
\newcommand{\bea}{\begin{eqnarray}}
\newcommand{\ena}{\end{eqnarray}}
\newcommand{\bean}{\begin{eqnarray*}}
\newcommand{\enan}{\end{eqnarray*}}
\newcommand{\bfig}{\begin{figure}}
\newcommand{\efig}{\end{figure}}
\newtheorem{proposition}{Proposition}
\newtheorem{lemma}[proposition]{Lemma}
\newtheorem{theorem}{Theorem}
\newtheorem{definition}{Definition}
\newtheorem{corollary}[proposition]{Corollary}
\newdimen\figcenter
\newdimen\texpscorrection
\def\putfigurewithtex#1 #2 #3 #4
{
{
\figcenter=\hsize\relax
\advance\figcenter by -#4truecm
\divide\figcenter by 2
\vskip #3truecm
\noindent\hskip\figcenter
\includegraphics{#1}
{\hskip\texpscorrection\input #2 }
}
}
\title{
Chaotic Properties of the Elliptical Stadium}

\author
{
Roberto Markarian\thanks{During this work, partially supported by Fac. de
Ciencias, Uruguay. }
\\
Inst. de Matem\'atica y Estad\'\i stica
''Prof. Ing. Rafael Laguardia'' \\
Fac. de Ingenier\'\i a, Uruguay.
\and
Sylvie Oliffson Kamphorst\thanks{Partially supported by CNPq, Brasil. }\\
S\^onia Pinto de Carvalho\\
Departamento de Matem\'atica, ICEx, UFMG, Brasil.
}

\date{}

\begin{document}
\maketitle

\begin{abstract}
The elliptical stadium is a curve constructed by joining two half-ellipses,
with half axes $a>1$ and $b=1$, by two straight segments of equal length $2h$.

Donnay \cite{kn:don} has shown that if $1<a<\sqrt 2$ and if $h$ is big enough,
then the corresponding billiard map has a positive Lyapunov exponent almost
everywhere; moreover, $h\rightarrow\infty$ as $a\rightarrow\sqrt 2$.

In this work we prove that if $1<a<\sqrt{4-2\sqrt 2}$, then
$h>2a^2\sqrt{a^2-1}$ assures the positiveness of a Lyapunov exponent.
And we conclude that, for these values of $a$ and $h$, the
elliptical stadium billiard mapping is ergodic and has the K-property.
\end{abstract}

\section{Introduction}

The plane billiard consists in the free motion of a point particle
on a connected bounded region in ${\cal R}^2$, being reflected elastically
at the boundary. The billiard defines a 2-dimensional discrete dynamical
system.

Depending on the boundary, this discrete dynamical system may have
different dynamical behaviour.
For instance, if it is a circle or an ellipse, it is known that
the system is integrable, and the phase space is ordered by
invariant curves (figure~\ref{fig:circ}).

A quite different situation appears when the components of the boundary have
negative curvature. The system is, then, ergodic and almost every orbit is
dense on the phase space (as in Sinai billiards).

The first example of an ergodic billiard with convex boundary was given by
Bunimovich \cite{kn:bun}. The boundary is the circular stadium, composed by
joining two half-circles by means of two straight segments of equal length.

A convex generalization is the elliptical stadium, composed by two
half-ellipses, with half axes $a\geq 1$ and $b=1$, joined by two straight
segments of equal length $2h$. For $a=1$, ie, for the circular stadium
billiard,
$h>0$ implies the existence of a positive Lyapunov exponent almost everywhere.
This will not be the case for $a\neq 1$ and small values
of $h$, because the structure of the phase space of the elliptical billiard
is not as degenerate as that of the circular billiard (figure~\ref{fig:circ}).

V. Donnay, in \cite{kn:don}, proved that for the billiard on the elliptical
stadium, if $1<a<\sqrt 2$ and $h$ is sufficiently big, then there is
a positive Lyapunov exponent almost everywhere. Also that $h$ must grow as
$a$ increases to $\sqrt 2$. And he gives a challenge: ''One could try to
calculate bounds on these lengths.''

In \cite{kn:can}, Canale and Markarian studied this problem numerically and,
varying $a$, found a value $h(a)$ such that if $0\leq h<h(a)$,
the billiard does not seem to be chaotic and, if $h>h(a)$, it seems ergodic.

In this article we give a partial answer to Donnay's challenge. We prove that
if
$1<a<\sqrt{4-2\sqrt 2}$, then $h>2a^2\sqrt{a^2-1}$ assures the positiveness of
a
Lyapunov exponent. To do so, we construct a cone-field, eventually strictly
invariant under the derivative of the billiard map, and apply Wojtkowski's
Theorem \cite{kn:woj}. The magic number $\sqrt{4-2\sqrt 2}$ is related to
the structure of bifurcation of  hyperbolic caustic periodic orbits of the
elliptical billiard (see figure~\ref{fig:bif}).

The existence of a positive Lyapunov exponent, meaning sensitive dependence on
initial conditions, does not automatically imply ergodicity. However, in this
example, as the numerical experiments suggest, ergodicity (and K-property)
can be proved. In fact, the additional conditions required on the Fundamental
Theorem of Sinai and Chernov are satisfied by this system (see \cite{kn:mark}
and \cite{kn:liv}).

\bfig
\vspace{4truecm}  \includegraphics{fig1.ps}
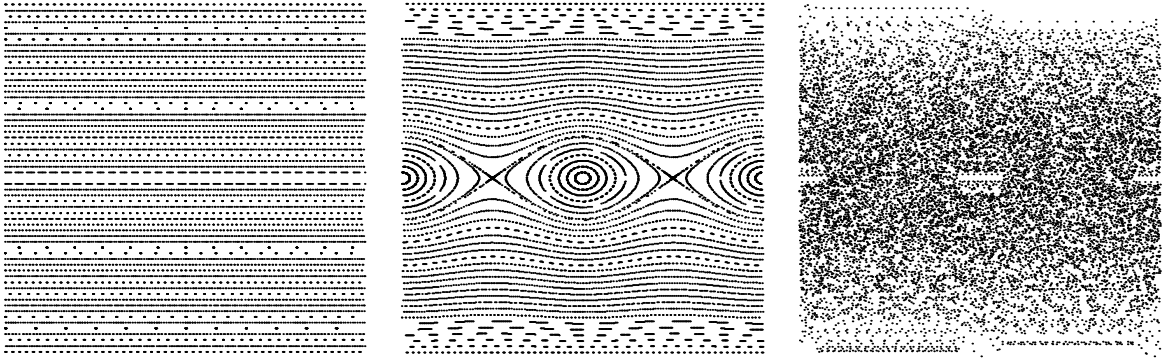
\caption{Phase space of the circular, the elliptical and the
         circular stadium billiards}
\label{fig:circ}
\efig

\section{Billiards}
Let $\Gamma$ be an oriented plane closed curve, or a finite set of
(topologically) closed curves on the plane, and $Q$ the region enclosed by
$\Gamma$. The billiard problem consists in the free motion of a point particle
on $Q$, being reflected elastically at the impacts on $\Gamma$. At any instant,
the state of the particle is determined by its position and
velocity. Since the motion is free on $Q$, all the motion is determined
either by two consecutive points of reflection at $\Gamma$ or by the point of
reflection and the direction of motion immediately after each reflection.

\bfig
\putfigurewithtex fig2.ps fig2.tex  4.5 16
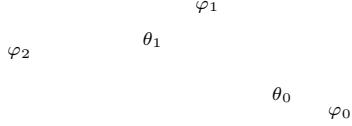
\caption{Coordinates for the elliptical billiard}
\label{fig:eli}
\efig

Let $t\in [0, L)$ be a parameter for $\Gamma$ and the direction of motion be
determined by the angle $\theta$ with the tangent to the boundary. The billiard
defines a map $T$ from the annulus ${\cal A}=[0, L) \times (0,\pi)$
into itself. Let $(t_0,\theta_0)$ and $(t_1,\theta_1) \in {\cal A}$ be such
that
$T(t_0,\theta_0)=(t_1,\theta_1)$. As usual, we will call $t$ either the
value of the parameter or the point $\Gamma(t)$ at the boundary.
If $\Gamma$ is $C^k, k \geq 2$, in some neighborhoods of $t_0$ and $t_1$, then
$T$ is a $C^{k-1}$-diffeomorphism in some neighborhoods of $(t_0,\theta_0)$ and
$(t_1,\theta_1)$ \cite{kn:str}. $T$ also preserves the measure
$d\mu = \sin \theta d\theta \frac{ds}{dt}dt$ \cite{kn:bir}, where $s$ is the
arclenght parameter for $\Gamma$.

$({\cal A},\mu, T)$ defines a discrete dynamical system, whose orbits are
given by $\{ (t_n,\theta _n)=T^n(t_0,\theta_0), n \in Z\}$.

The polygonal on $Q$, whose vertices are $\Gamma (t_n), n \in Z $, is often
called the trajectory of the particle. Because of the obvious correspondence
between orbits and trajectories, we will call both of them either orbits or
trajectories.

\subsection{The Elliptical Billiard}

Given an orthogonal coordinate system in ${\cal R}^2$, let $\Gamma$ be the
ellipse $\frac{x^2}{a^2}+y^2=1$, with $a>1$. We will use the angle $\varphi$
between the oriented tangent and the $x$-axis as a parameter for $\Gamma$
(figure~\ref{fig:eli}). We will call half-ellipses the pieces of $\Gamma$
corresponding to $0\leq \varphi \leq \pi$ and $\pi \leq \varphi \leq 2\pi$.

Each trajectory has a conic caustic, confocal with $\Gamma$. If a segment of a
given trajectory cuts the segment joining the two foci, all the other segments
of this trajectory will cut it and the caustic will be a hyperbola. If it
passes by  one focus, the trajectory will always pass by the foci, thus having
the two foci as a degenerate caustic. Otherwise, the caustic will be an ellipse
(see, for instance \cite{kn:cha}).

The phase space of the elliptical billiard is the annulus
${\cal A}=[0,2\pi)\times (0,\pi)$, with coordinates $\varphi$ and $\theta$.
As $\Gamma$ is $C^{\infty}$, the map $T:{\cal A} \rightarrow {\cal A}$ is a
$C^{\infty}$-diffeomorphism, where
$(\varphi_1,\theta_1)=T(\varphi_0,\theta_0)$ is given by:
\bea
a\tan \varphi_1 &=&\frac{a\tan \gamma -a\tan \varphi_0}
{1+(a\tan \gamma )(a\tan \varphi_0)} \ ,
\  \ a\tan \gamma = \frac{2a\tan (\varphi_0+\theta_0)}
{1-(a\tan (\varphi_0+\theta_0))^2} \nonumber \\
\theta_1&=& \varphi_1 -(\varphi_0+\theta_0)
\ena
and the derivative of $T$ is (see, for instance \cite{kn:str}):
\be
DT_{(\varphi_0,\theta_0)}=\frac{1}{R_1\sin \theta_1}\left(
\begin{array}{cc}
l_{01}-R_0\sin\theta_0&l_{01}\\
l_{01}-R_0\sin\theta_0-R_1\sin\theta_1&l_{01}-R_1\sin\theta_1
\end{array}\right),
\label{eq:derivelipse}
\en
where $l_{01}$ is the distance between $(x(\varphi_0), y(\varphi_0))$ and
$(x(\varphi_1), y(\varphi_1))$, and
$R_{i}=a^2/(a^2\sin^2 \varphi_{i}+\cos^2 \varphi_{i})^{3/2}$
is the radius of curvature at $\varphi_{i}$, $i=0, 1$.

$T$ has no fixed points and has exactly four two-periodic orbits, being
two elliptic
(both eigenvalues of $DT^2$ have moduli one) and two hyperbolic (the moduli
of the eigenvalues of $DT^2$ are one smaller and the other greater than one).

The function $F(\varphi,\theta)=\frac{\cos
^2\theta-\varepsilon^2\cos^2\varphi}{1-\varepsilon^2\cos^2\varphi}$, where
$\varepsilon=\frac{\sqrt{a^2-1}}{a}$ is the eccentricity of the ellipse
$\Gamma$, is a first integral
for $T$, ie, $F$ is constant on orbits under $T$ (see, for instance,
\cite{kn:lev}). Physically, $F$ may be interpreted as the product of angular
momenta about the two foci \cite{kn:berry2}. Geometrically,
$\sqrt{\mid F(\varphi_0,\theta_0)\mid}$ measures the length of the minor axis
of the conical caustic \cite{kn:fel}.
So, for every $a>1$, the phase space ${\cal A}$ is foliated by
the level curves of $F$ (figure~\ref{fig:circ}).

Let us call $\cal E$=\{ trajectories with elliptical caustic\} and
$\cal H$=\{ trajectories with hyperbolic caustic\}.
The obvious correspondence between trajectories on the configuration space
$Q$ and orbits on the phase space ${\cal A}$ gives  the following geometrical
interpretation:
\begin{itemize}
\item trajectories over the major axis $\longleftrightarrow$ period 2,
hyperbolic
orbits , $F=0$
\item trajectories over the minor axis $\longleftrightarrow$ period 2, elliptic
orbits, $F=1-a^2$
\item trajectories that pass by the foci $\longleftrightarrow$ saddle
connection,
$F=0$
\item trajectories on ${\cal E}$ $\longleftrightarrow$ orbits with $F>0$
\item trajectories on ${\cal H}$ $\longleftrightarrow$ orbits with $1-a^2<F<0$.
\end{itemize}
Because of this interpretation, we will also call $\cal E$ the
set of orbits with $F>0$ and $\cal H$ the set of orbits with $1-a^2<F<0$.

Poncelet's theorem \cite{kn:cha} assures that trajectories that
share the same caustic have the same dynamical behaviour.
And to each conical caustic, or equivalently, to each integral curve $F=k$, we
can associate a rotation number (see, for instance,\cite{kn:pal}).

\begin{itemize}
\item
For $k>0$ (trajectories on ${\cal E}$) Jacobi coordinates (see,
for instance \cite{kn:cha}) give a natural conjugation between ${\cal E}$ and
the trajectories of the circular billiard with $\theta \neq \pi /2$. It
follows that, to each $k>0$ we can associate a rotation number
$\rho(k) \in (0,1)$, such that:
\begin{itemize}
 \item $\rho = \frac{n}{p} \in {\cal Q}, (n,p)\equiv 1$ corresponds to
    periodic orbits of period $p$,
    with $T^p(\varphi_0, \theta_0) = $ \break $(\varphi_0 + 2n\pi, \alpha_0)
\cong
    (\varphi_0 , \alpha_0)$.
 \item $\rho \in {\cal R} \setminus {\cal Q}$ corresponds to dense orbits
  (dense on the level curve).
\end{itemize}

\item
For $1-a^2<k<0$ (trajectories on ${\cal H}$) notice
that every periodic orbit has an even period and $(0,\pi/2)$ and
$(\pi,\pi/2)$ are elliptic fixed points of the integrable, measure
preserving diffeomorphism $T^2$, with
eigenvalues
$$\lambda_j =e^{2\pi i\nu_j (a)},\:\hbox{where}\:
\nu_j(a)= \frac{1}{\pi}
          \arctan \frac{\pm 2 \sqrt{a^2-1}}{2-a^2},\,\,\,
j=1,2.$$
Notice that for $a=\sqrt 2$, $\lambda_j=\pm 1$.

So, to every level curve it is associated a rotation number $\tau$
such that
$$ \nu (a)=\frac{1}{\pi}
            \arctan \frac{2 \sqrt{a^2-1}}{a^2-2}<\tau <1.$$
\begin{itemize}
  \item  $\tau=\frac{n}{p} \in {\cal Q}, (n,p)\equiv 1$, corresponds to
    periodic orbits of period $2p$, crossing $2n$ times the minor axis of
    $\Gamma$.
 \item $\tau \in {\cal R} \setminus {\cal Q}$ corresponds to dense orbits
  (dense on the level curve).
\end{itemize}
\end{itemize}

Notice that $\nu (a)$ is strictly decreasing, $\nu (a) \longrightarrow 1$,
as $a \longrightarrow 1^+$ and $\nu (a) \longrightarrow 0^+$, as $a
\longrightarrow +\infty$.
This indicates the first difference between the trajectories on ${\cal E}$ and
on ${\cal H}$: for each $a>1$, we have periodic orbits of every period
on ${\cal E}$, but not on ${\cal H}$. For instance, orbits of period 4
exist only for $a>\sqrt{2}$, or there is no period 8 orbits if
$a<\sqrt{4-2\sqrt 2}$. On figure~\ref{fig:bif}, we show the bifurcation
diagram for periodic orbits on ${\cal H}$.

\bfig
\putfigurewithtex fig3.ps fig3.tex  8 16
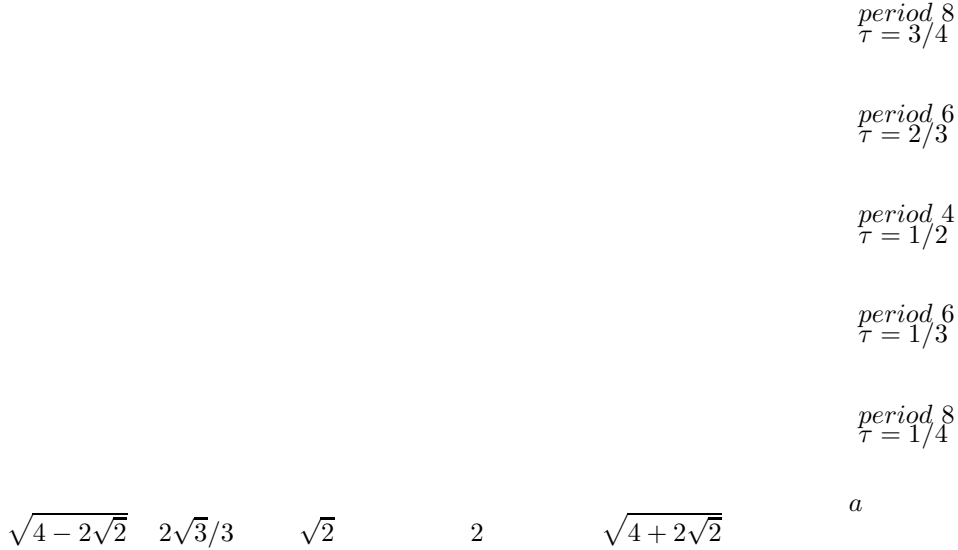
\caption{Diagram of bifurcation of periodic orbits on ${\cal H}$}
 \label{fig:bif}
\efig

Another difference lies on where the trajectory touches the conical caustic.
The tangency to the caustic on $\cal E$ always occurs inside $Q$ (ie, inside
the
original ellipse) and so, inside the segment of trajectory that joins two
reflections at $\Gamma$. However, on $\cal H$, the tangency may occur outside
$Q$, ie, you do not necessarily have the point of tangency between two
reflections at $\Gamma$. It may occurs even at infinity, if a segment of
trajectory is contained on an asymptote of the caustic hyperbola.

The results below give conditions to allow tangencies on $Q$ for trajectories
on ${\cal H}$.
\begin{lemma} \label{lem:D}
Suppose that a trajectory $\{T^n(\varphi_0,\theta_0)\}\in{\cal H}$ has two
consecutive reflections at the same half-ellipse. If $1<a<\sqrt 2$, then the
tangency between the segment joining those two reflections and the hyperbolic
caustic occurs inside the billiard.
\end{lemma}
proof:
Suppose that those two consecutive reflections are at $\varphi_0\in
[0,\pi/2]$ and $\varphi_1\in [\pi/2,\pi]$ (the other cases are symmetrical).
We will call $\overline{\varphi_0\varphi_1}$ the segment joining
$\Gamma(\varphi_0)$ to $\Gamma(\varphi_1)$.

For a fixed $\varphi_0$, if $\theta_0=\pi/2-\varphi_0$, then
$\overline{\varphi_0\varphi_1}$ touches the caustic at its vertex (and so,
inside $Q$).
Suppose that there is a $\theta$ such that $\varphi_1\in [\pi/2,\pi]$, and
$\overline{\varphi_0\varphi_1}$ touches the caustic outside $Q$. Then there is
a $\tilde\theta_0$ such that $\varphi_1\in [\pi/2,\pi]$, and
$\overline{\varphi_0\varphi_1}$ has the tangency exactly at the boundary.
But this means that either $\Gamma(\varphi_0)$ or $\Gamma(\varphi_1)$ is the
intersection of the hyperbolic caustic and the elliptical boundary and, since
they form an orthogonal family, either $\tilde\theta_0$ or $\theta_1 = \pi/2$.

But if $1<a<\sqrt 2$, $\varphi_0\in [k\pi/2,(k+1)\pi/2]$ and $\theta_0=\pi/2$,
then $\varphi_1\in [(k+2)\pi/2,(k+3)\pi/2]$ (\cite{kn:don}), and this implies
that $\varphi_0$ and $\varphi_1$ are not on the same half-ellipse.

\begin{lemma} \label{lem:DD}
Suppose that a trajectory $\{T^n(\varphi_0,\theta_0)\}\in{\cal H}$ has two
consecutive reflections, $\varphi_0$ and $\varphi_1$, at the same half-ellipse.
If $1<a<\sqrt{4-2\sqrt 2}$, then the tangency between each one of the segments
$\overline{\varphi_{-1}\varphi_0}$ and $\overline{\varphi_0\varphi_1}$
and the hyperbolic caustic occurs inside the billiard.
\end{lemma}
proof:
$\{T^n(\varphi_0,\theta_0)\}\in{\cal H}$. So,
$F(\varphi_0,\theta_0)=k_0<0$. Fix this hyperbolic caustic $k_0$. Call
$\theta(\varphi)$ the solution of $F(\varphi,\theta(\varphi))=k_0$. The piece
of trajectory $(\varphi_{-1},\theta(\varphi_{-1}))$, $(0,\theta(0))$,
$(\varphi_1,\theta(\varphi_1))$ is such that $0$ and $\varphi_1\in [0,\pi]$,
and $\varphi_{-1}$ and $0\in [\pi,2\pi]$. So, by lemma~\ref{lem:D}, and since
$\sqrt{4-2\sqrt 2}<\sqrt 2$, $\overline{0\varphi_1}$ and
$\overline{\varphi_{-1}0}$ touch the hyperbola inside $Q$.
Suppose that for a given $\varphi_0\in [0,\pi/2]$,
$F(\varphi_0,\theta_0)=k_0$, $\varphi_1\in [\pi/2,\pi]$
and $\overline{\varphi_{-1}\varphi_0}$
touches the hyperbola outside $Q$. Then, there exists a
$\tilde{\varphi_0}\in [0,\pi/2]$, with
$F(\tilde{\varphi_0},\theta(\tilde{\varphi_0}))=k_0$,
$\varphi_1\in [\pi/2,\pi]$, and $\overline{\varphi_{-1}\tilde{\varphi_0}}$
has the tangency exactly at the boundary. Since $\varphi_1\in [\pi/2,\pi]$ and
$a<\sqrt 2$, $\theta(\varphi_{-1})=\pi/2$.

Fix now the configuration $(\varphi_{-1},\pi/2)\mapsto
(\varphi_0,\theta_0)\mapsto (\varphi_1,\theta_1)$, $\varphi_{-1}\in
[3\pi/2,\pi]$, $\varphi_0\in [0,\pi/2]$, $\varphi_1\in [\pi/2,\pi]$.
There exists a $k_1$, $0>k_1>k_0$, such that $F(\varphi_0,\theta_0)=k_1$, and
the piece of trajectory is $(\varphi_{-1},\pi/2)\mapsto
(\varphi_0,\theta_0)\mapsto (\pi,\theta_1)$. Completing this trajectory, we
see that it has period 8 and crosses the minor axis 6 times,
or has $\tau =3/4$ that exists only if $a>\sqrt{4-2\sqrt 2}$.

\begin{corollary} \label{coro:DDD}
Suppose that a trajectory $\{T^n(\varphi_0,\theta_0)\}\in{\cal H}$ has two
consecutive reflections $\varphi_0$ and $\varphi_1$ at the same half-ellipse.
If $1<a<\sqrt{4-2\sqrt 2}$, then the tangency between each one of the segments
$\overline{\varphi_{-1}\varphi_0}$, $\overline{\varphi_0\varphi_1}$ and
$\overline{\varphi_1\varphi_2}$ and the hyperbolic caustic occurs inside the
billiard.
\end{corollary}
proof: Apply lemma~\ref{lem:DD} twice.

\subsection{ The Billiard on the Elliptical Stadium}
\label{estadio}

The boundary of the elliptical stadium is a curve $\Gamma$ constructed by
joining two half-ellipses, with major axes $a>1$ and minor axes $b=1$, by two
segments of equal length $2h$, as shown in figure~\ref{fig:estadio}.
\bfig
\putfigurewithtex fig4.ps fig4.tex 2.5 16
 \caption{Elliptical stadium}
 \label{fig:estadio}
\efig

$\Gamma$ is a Jordan curve, convex, but not strictly convex, globally $C^1$
but not $C^2$, and piecewise $C^{\infty}$. We parametrize $\Gamma$ by the angle
$\varphi$ on the half-ellipses and by the arclenght parameter on the straight
parts.

The billiard map associated to $\Gamma$
$$
\begin{array}{cccc}
S:&[0,2\pi+4h]\times (0,\pi)&\longrightarrow&[0,2\pi+4h]\times (0,\pi)\\
&(t_0,\theta_0)&\longmapsto&(t_1,\theta_1)
\end{array}$$
is piecewise $C^{\infty}$, but globally only $C^0$.

The derivative of $S$ is: if $t_0$ and $t_1$ belong to the elliptical part,
then $DS_{(t_0,\theta_0)}$ is equal to (\ref{eq:derivelipse});\\
if $t_0$ belongs to the elliptical part and $t_1$ to the straight part, then
\be
DS_{(t_0,\theta_0)}=\frac{1}{\sin \theta_1}\left(
\begin{array}{cc}
k_0 l_{01}-\sin\theta_0 & l_{01}\\
-k_0\sin\theta_1 & -\sin\theta_1
\end{array}\right);
\label{eq:derivestadio,er}
\en
if $t_0$ belongs to the straight part and $t_1$ belongs to the elliptical part,
then
\be
DS_{(t_0,\theta_0)}=\frac{1}{\sin \theta_1}\left(
\begin{array}{cc}
-\sin\theta_0 & l_{01}\\
-k_1\sin\theta_1 & k_1 l_{01}-\sin\theta_1
\end{array}\right);
\label{eq:derivestadio,re}
\en
and if $t_0$ and $t_1$ belong to the straight part, then
\be
DS_{(t_0,\theta_0)}=\left(
\begin{array}{cc}
-1 & l_{01}/\sin\theta_1\\
 0 & -1
\end{array}\right),
\label{eq:derivestadio,rr}
\en
where $k_i$ is the curvature of the ellipse at $\varphi_i$ and $l_{01}$ is the
length of the segment $\overline{t_0t_1}$.

It is easy to see that for every $(t_0,\theta_0)$, with $t_0\in (0,\pi)$
(resp. $(\pi +2h, 2\pi +2h)$), there exists an $n=n(t_0,\theta_0)$ such that
$S^n(t_0,\theta_0)=(t_n,\theta_n)$, with $t_n \in (\pi +2h, 2\pi +2h)$
(resp. $(0,\pi)$), ie, if we begin in a half-ellipse, we will reach the other
one on a finite number of iterations.

This allows us to construct a restricted map:
$$
\begin{array}{cccc}
 R:&{\cal B}&\longrightarrow&{\cal B} \\
&(\varphi_0,\theta_0)&\longmapsto&(\varphi_1,\theta_1)
\end{array}
$$
where ${\cal B}=\{[0,\pi ]\bigcup \, [\pi+2h, 2\pi+2h ]\} \times (0,\pi)$,
$\varphi_1$ is the next reflection at a half-ellipse and $\theta_1$ is the
direction of motion immediately after the reflection at $\varphi_1$.

Notice that while the reflections remain at the same half-ellipse, $R$ is just
$T$, the diffeomorphism associated to the elliptical billiard. So, for those
reflections, the orbit will stay on an integral curve of the elliptical
billiard.
\bfig
\putfigurewithtex fig5.ps fig5.tex 4 16
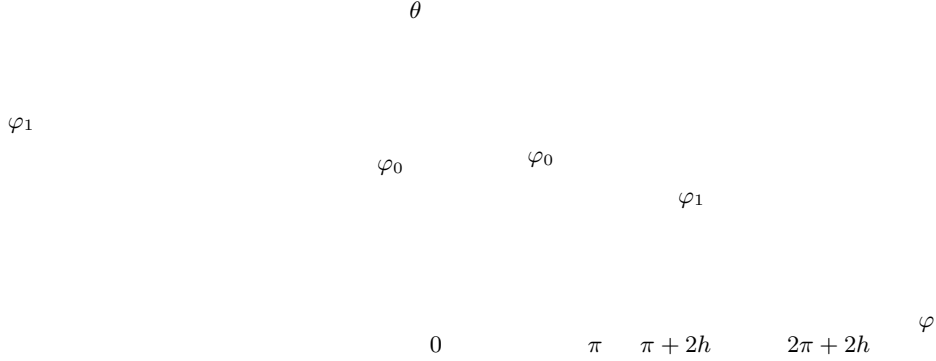
\caption{Integral curves and caustics in the half-ellipses}
 \label{fig:meia}
\efig

\section{Main Tools}

\subsection{Cone-fields and consequences}
\label{cones}

Let $\Gamma =\{\gamma(t)=(x(t), y(t)), t\in[0, L)\}$ be a $C^k$-piecewise
Jordan
curve in $R^2$, $k\geq 1$, $Q$ the region of $R^2$ bounded by $\Gamma$ and
${\cal A}$ the annulus $[0, L)\times (0,\pi)$.

Let $T:{\cal A}\rightarrow{\cal A}$ be the local $C^{k-1}$ measure preserving
diffeomorphism associated to the billiard on $Q$.

\begin{definition}
For each $(t,\theta)\in{\cal A}$, the Lyapunov exponents of $T$ at
$(t,\theta)$ are given by:
\bean
\lambda_{+}(t,\theta)&=&\lim_{n\rightarrow \infty}\frac{1}{n} \ln
\parallel DT^n(t,\theta)\parallel \nonumber \\
\lambda_{-}(t,\theta)&=&-\lambda_{+}(t,\theta)
\enan
\end{definition}

\begin{definition}
Given
two linearly independent vector fields $X_1(t,\theta)$ and
$X_2(t,\theta)$ in the tangent space $T_{(t,\theta)}{\cal A}\approx R^2$,
a {\bf cone} in $T_{(t,\theta)}{\cal A}$ is defined by
$$C(t,\theta)=\{rX_1(t,\theta)+sX_2(t,\theta), r. s\geq 0\}. $$
Its interior is
$$int(C(t,\theta))=\{rX_1(t,\theta)+sX_2(t,\theta), r.s>0 \:\hbox{or}\:\:
r=s=0\}. $$
A {\bf measurable cone-field} is a family of cones $\{C(t,\theta)\}\subset TA$
defined $\mu$-almost everywhere, and such that the vectors $X_1(t,\theta)$ and
$X_2(t,\theta)$ vary measurably with $(t,\theta)$.
\end{definition}

{\bf THEOREM} (Wojtkowski,\cite{kn:woj})
{\em
Let $C(t,\theta)$ be a
measurable cone field such that for almost every $(t,\theta)$,
\samepage
$$DT(C(t,\theta))\subset C(T(t,\theta))$$
\samepage
and for almost every $(t,\theta)$ there exists a $k(t,\theta)$ for which
\samepage
$$DT^{k(t,\theta)}(C(t,\theta))\subset intC(T^{k(t,\theta)}(t,\theta)). $$
\samepage
Then the Lyapunov exponents $\lambda_{+}(t,\theta)$ are positive for almost
every $(t,\theta)$.
}

(Such a cone-field is called eventually strictly invariant.)

Our goal is to construct a measurable cone-field, eventually strictly
invariant under the derivative $DS$ of the map associated to the billiard on
the elliptical stadium. But $S$ is not differentiable at
$t=0$, $\pi$, $\pi +2h$, $2\pi +2h$.

For $\alpha \in [0,2\pi+4h]$, let
\begin{eqnarray}
W_{\alpha}&=&\{(t_0,\theta_0)\in [0,2\pi+4h]\times (0,\pi)\:
\hbox{such that}\:
\exists n\in Z,\: S^n(t_0,\theta_0)=(\alpha,\theta_n)\} \nonumber \\
&=&\bigcup_{n\in Z}S^n(\{(\alpha,\theta),\theta\in (0,\pi)\}) \nonumber
\end{eqnarray}
ie, the images of a vertical line on $[0,2\pi+4h]\times (0,\pi)$.
$\mu (W_{\alpha})=0$ because $\mu (\{(\alpha,\theta),\theta \in (0,\pi)\})=0$
and $S$ is measure preserving. Let $W=W_0\bigcup W_{\pi}\bigcup
W_{\pi+2h}\bigcup W_{2\pi +2h}$. Then $\mu (W)=0$ and we will work on
$[0,2\pi+4h]\times (0,\pi)\setminus W$.

On the other hand, on the straight part of the elliptical stadium, $DS$ is
just a translation and an inversion. So, it is enough to construct
a cone-field strictly invariant under $DR$, the derivative of the restricted
map $R$, on ${\cal B}\setminus (W\bigcap {\cal B})$.

Finally, if a given trajectory, between $(\varphi_0,\theta_0)$ and
$(\varphi_1,\theta_1)=R(\varphi_0,\theta_0)$ hits $q$ times the straight part
of the elliptical stadium, then

\be
DR_{(\varphi_0,\theta_0)}=\frac{(-1)^q}{R_1\sin \theta_1}\left(
\begin{array}{cc}
l_{01}-R_0\sin\theta_0 & l_{01} \\
l_{01}-R_0\sin\theta_0-R_1\sin\theta_1& l_{01}-R_1\sin\theta_1
\end{array}\right).
\en
So, the image under $DR$ of a cone $C(\varphi_0,\theta_0)$ can be inverted
when arriving at $(\varphi_1,\theta_1)$.
But, as
$-C=\{r(-X_1)+s(-X_2),\: rs\geq 0\}=\{rX_1+sX_2,\: rs\geq 0\}=C$, all that will
interest us is the slope of the boundaries of the cone.

\subsection{A little bit of optics}

We are going to follow Donnay ideas \cite{kn:don}, giving an optical
interpretation of the tangent space.

{}From now on, we are going to suppose that $\Gamma$ is strictly convex
(although the elliptical stadium is not strictly convex, we are going
to reduce our problem to the elliptical part of the boundary, looking only to
the restricted map $R$).

\begin{definition}
A {\bf pencil of rays} is a $C^1$ curve in the phase space ${\cal A}$,
given by $\eta:(-\epsilon,\epsilon)\longrightarrow {\cal A}$,
$\eta (\sigma)=(\varphi(\sigma),\theta(\sigma))$.
The {\bf base point} of the pencil is $\eta(0)=(\varphi_0,\theta_0)$, the {\bf
span} is $\eta '(0)=(\alpha,\beta)\in T_{(\varphi_0,\theta_0)}{\cal A}$ and the
{\bf slope} is $\beta/\alpha$.

To each $\sigma\in (-\epsilon,\epsilon)$, we can associate a straight line
$r(\sigma)$, passing by $\Gamma(\varphi(\sigma))$, with slope
$\tan(\varphi(\sigma)+\theta(\sigma))$. The {\bf core} of the pencil is the
straight line $r_0=r(0)$. Let $P(\sigma)=r(\sigma)\bigcap r_0,\: \sigma\neq
0$. A pencil $\eta$ {\bf focus} at $P_0=\lim_{\sigma \rightarrow 0}
P(\sigma)$, if the limit exists.
\end{definition}

If $P_0=(x,y)$ exists, it is given by the equations
\bean
\sec^2(\varphi_0+\theta_0)(1+\frac{\beta}{\alpha})(x-x_0)&
=&(\tan(\varphi_0+\theta_0)
 \cos\varphi_0-\sin\varphi_0)R_0 \nonumber \\
y-y_0&=&\tan(\varphi_0+\theta_0)(x-x_0),
\enan
 where $(x_0, y_0)=\Gamma(\varphi_0),\: \theta_0=\theta(0)$ and $R_0$ is the
radius of curvature of $\Gamma$ at $\varphi_0$.

So $P_0$ depends only on the base point $(\varphi_0,\theta_0)$ and the slope
$\beta/\alpha$.
{}From the point of view of focusing properties, we may identify
pencils with the same base point and slope to either the linear pencil
$\eta(\sigma)=(\varphi_0+\alpha\sigma,\theta_0+\beta\sigma)$ or to the point
$(\varphi_0,\theta_0,\alpha,\beta)\in T{\cal A}$. We can also define and
calculate the focusing distance
$$f=dist(P_0, (x_0, y_0))=\frac{R_0\sin\theta_0}{1+\frac{\beta}{\alpha}}.$$

Notice that $R_0\sin\theta_0$ is the length of the intersection of the core of
the pencil with the half-osculating circle at $\varphi_0$ (the half-osculating
circle is the circle with radius $R_0/2$, tangent to $\Gamma$ at $\varphi_0$).

\subsection{Pencils of rays and beams of trajectories}

When we give a pencil of rays, we give a collection of initial conditions of
the billiard ball problem inside $\Gamma$. So, to a pencil of rays there
corresponds a beam of trajectories.

We may have to check if a given beam of trajectories
$\eta(\sigma)=(\varphi(\sigma),\theta(\sigma))
\approx (\varphi_0+\alpha\sigma,\theta_0+\beta\sigma)$ focuses forward and
backward, after and before the reflections at $\Gamma(\varphi(\sigma))$.
The forward focusing distance is given by
\be
f_s=\frac{R_0\sin\theta_0}{1+\frac{\beta}{\alpha}}
\label{eq:fs}
\en
and the backward,
\be
f_e=\frac{R_0\sin\theta_0}{1-\frac{\beta}{\alpha}}
\label{eq:fe}
\en
because going backward is the same as going forward for the pencil
$$\overline{\eta}(\sigma)=(\varphi(\sigma),\pi-\theta(\sigma))
\approx
(\varphi_0+\alpha\sigma,\pi-(\theta_0+\beta\sigma))
=(\varphi_0+\alpha\sigma,\pi-\theta_0-\beta\sigma)\ . $$

To compare the image of a pencil with another pencil, we will use Donnay's
Focusing Lemma \cite{kn:don}, rewritten as:

{\bf FOCUSING LEMMA}
{\em Let $T:{\cal A}\rightarrow
{\cal A}$ be the map associated to the billiard problem at $\Gamma$. Let
$(\varphi_0,\theta_0)$, $(\varphi_1,\theta_1)$ $\in{\cal A}$,
$(\varphi_1,\theta_1)=T(\varphi_0,\theta_0)$.
Let $\eta(\sigma)$ be a pencil with base point $(\varphi_0,\theta_0)$,
focusing forward at $P_0$ and $\xi(\sigma)$, a pencil with base point
$(\varphi_1,\theta_1)$, focusing backward at $P_1$. If
$P_0$ appears before $P_1$ when going from
$\Gamma(\varphi_0)$ to $\Gamma(\varphi_1)$ along the trajectory, then
$$\mbox{slope}(T\circ\eta(\sigma))>\mbox{slope}(\xi(\sigma)). $$}

\section{Construction of a cone-field on TB}

\subsection{Zones of the phase space B} \label{zonasB}

Take $R:{\cal B}\rightarrow {\cal B}$, the restricted map defined in
subsection~\ref{estadio}.
${\cal B}={\cal B}_1\bigcup{\cal B}_2$,
${\cal B}_1=[0,\pi] \times (0,\pi)$,
${\cal B}_2=[\pi+2h, 2\pi+2h]\times (0,\pi)
\approx [\pi, 2\pi]\times (0,\pi)$,
each one corresponding to the reflections at one half-ellipse.
Let $B={\cal B}\setminus (W\bigcap {\cal B})$, $W$
the measure zero set defined in subsection~\ref{cones}.
$B=B_1 \bigcup B_2$,
$B_i={\cal B}_i\setminus (W\bigcap {\cal B}_i),\:i=1, 2. $

Let  $B_1=U_1\bigcup \, M_1 $ where
\bean
U_1 &=&
\{ (\varphi_0,\theta_0)\in  B_1
\hbox{ such that } R(\varphi_0,\theta_0) \in B_2 \hbox{ and }
R^{-1}(\varphi_0,\theta_0) \in B_2\}\\
M_1 &=&
\{ (\varphi_0,\theta_0)\in
B_1 \: \hbox{such that}\:\hbox{either}\: R(\varphi_0,\theta_0) \in
B_1 \: \hbox{or} \:
R^{-1}(\varphi_0,\theta_0) \in B_1\}
\enan

The boundary of $U_1$ is composed by two curves
$\varphi\mapsto\theta_{\pm}(\varphi)$ such that
$R(\varphi,\theta_{+}(\varphi))=(\pi,\theta_1)$ and
$R^{-1}(\varphi,\theta_{-}(\varphi))=(0,\theta_1)$
with equation:
$$\tan(\varphi+\theta_{\pm}(\varphi))=\frac{\tan\varphi}
{1\mp\sqrt{1+a^2\tan\varphi}}. $$
Notice that this boundary is contained on $W$.

By symmetry, we define $U_2$ and $M_2$, with $B_2=U_2 \bigcup\, M_2. $

\subsection{Caustic, Vertical and Horizontal Pencils}
\label{sec:caus}

If $(\varphi_0,\theta_0) \in M_1$ (resp. $M_2$), there exist $m, n \in Z,\, m
\leq 0\leq n, m\neq n$ such that $R^{m}(\varphi_0,\theta_0)$,
$R^{m+1}(\varphi_0,\theta_0)$, $\ldots$, $R^{n-1}(\varphi_0,\theta_0)$, $
R^{n}(\varphi_0,\theta_0) \in M_1$ (resp. $M_2$) \cite{kn:don}.
And $R^{i}=T^{i}, m\leq i\leq n$, because $M_1\subset B_1$
(resp. $M_2\subset B_2$). So, there is a piece
of the orbit of $(\varphi_0,\theta_0)$ that behaves exactly as it would
on the elliptical billiard. It stays on an integral curve $F=k$ of the
elliptical billiard and this piece of trajectory  has a caustic.

\begin{definition}
The caustic pencil at $(\varphi_0,\theta_0)$ is a curve
$\eta:(-\epsilon,\epsilon)\rightarrow B_1$,
$\eta(\sigma)=(\varphi(\sigma),\theta(\sigma))$
and $F(\eta(\sigma))=F(\varphi_0,\theta_0)$
or, by identification, a tangent vector at $(\varphi_0,\theta_0)$ to the
integral curve $F=F(\varphi_0,\theta_0)$.
\end{definition}

Clearly, a caustic pencil focuses forward and backward at the point of tangency
of its core with the conic caustic. Also, caustic pencils are transformed by
$T$ on caustic pencils.

\begin{lemma} \label{coro:tangelipse}
On the elliptical billiard, with $a<\sqrt 2$, choose a trajectory on ${\cal
E}$ such that $\varphi_0\in [\pi/2,\pi]$, $\varphi_0+\theta_0\geq \pi$ and
$\varphi_1\in[\pi, 2\pi]$. The point $P_0$ of tangency of
$\overline{\varphi_0\varphi_1}$ with the elliptical caustic is contained in
the interior of the half-osculating circle at $\varphi=\pi$.
\end{lemma}
proof: Let us fix one elliptical caustic. To each $(\varphi,\theta)$,
$\varphi\in [\pi/2,\pi]$, $\varphi+\theta\geq\pi$, we associate $P(\varphi)$,
the point of tangency with the caustic. Let $\tilde{\varphi_0}$ be such that
$P(\tilde{\varphi_0})$ is the upper vertex of the elliptical caustic.
Ordering the elliptical caustic via the counter clockwise orientation, we
have that $P(\tilde{\varphi_0})\leq P(\varphi)\leq P(\pi)$.
It is clear that $P(\tilde{\varphi_0})$ is contained in the interior of the
half-osculating circle at $\varphi=\pi$.
The integral curve associated to this caustic has a minimum at $\varphi=\pi$.
So the maximum of the forward focusing distance for the caustic pencil is
$f_{s}=\frac{R_{\pi}\sin\theta}{1+0}=a^2\sin\theta$.

\begin{definition}
The vertical pencil is defined by
$v(\sigma)=(\varphi_0,\theta_0+\sigma)$ or by the vector $(0, 1) \in
T_{(\varphi_0,\theta_0)}B$.
\end{definition}

The vertical pencil $v(\sigma)$ focuses forward and backward at
$\Gamma(\varphi_0)$, ie, at the boundary of the billiard, since $\alpha=0,\:
\beta=1$ and $f_{s, e}= \alpha\frac{R_0\sin\theta_0}{\alpha\pm \beta}=0$.

\begin{lemma} \label{pro:vert}
Let $a<\sqrt{2}$, $h>0$ and $(\varphi_1,\theta_1)=R(\varphi_0,\theta_0)$. Let
$v_0(\sigma)$ and $v_1(\sigma)$ be the vertical pencils at
$(\varphi_0,\theta_0)$ and $(\varphi_1,\theta_1)$, respectively. Let
$\eta(\sigma)=(\varphi_1+\alpha\sigma,\theta_1+\beta\sigma)$ be the
linearization of the pencil $R(v_0(\sigma))$. Then $\alpha\neq 0$ and
$\frac{\beta}{\alpha}>0$.
\end{lemma}
proof:
$(\alpha,\beta)=DR_{(\varphi_0,\theta_0)}(0, 1)=\frac{(-1)^n}{R_1\sin\theta_1}
(l_{01}, l_{01}-R_1\sin\theta_1). $
So, $\alpha\neq 0$ and $\frac{\beta}{\alpha}=
\frac{l_{01}-R_1\sin\theta_1}{l_{01}}$.

If $a<\sqrt 2$, the half-osculating circles of an ellipse with half axes $a$
and 1 are totally contained in the ellipse \cite{kn:don}.
So, $l_{01}>R_1\sin\theta_1$ and $\frac{\beta}{\alpha}>0$.

Notice that, geometrically, this lemma shows that $R$ deviates the vertical to
the right.

\begin{definition}
The horizontal pencil is defined by $h(\sigma)=(\varphi_0+\sigma,\theta_0)$
or by the vector $(1, 0)\in T_{(\varphi_0,\theta_0)}B$.
\end{definition}

The horizontal pencil focuses forward and backward at a distance
$R_0\sin\theta_0$, ie, at the intersection of the core with the half-osculating
circle at $\varphi_0$.

\subsection{ A cone field on TB} \label{coneB}

\begin{definition}
For each $(\varphi,\theta)\in B$, define the measurable cone field
$\{C(\varphi,\theta)\}\subset TB$ by
\begin{itemize}
\item
if $(\varphi,\theta)\in U_1$ or $U_2$, $C(\varphi,\theta)=\{rX_v+sX_h,\,
r. s\geq 0\}$,
\item if $(\varphi,\theta)\in M_1$ or $M_2$, $C(\varphi,\theta)=\{rX_v+sX_c,\,
r. s\geq 0\}$,
\end{itemize}
where $X_h =(1, 0)$ is the horizontal pencil at $(\varphi,\theta)$,
$X_v =(0, 1)$ is the vertical pencil at $(\varphi,\theta)$ and
$X_c =(1, -\frac{\partial F/\partial \varphi}{\partial F/\partial \theta})$,
where $F$ is the first integral of the elliptical billiard, is the caustic
pencil at $(\varphi,\theta)$.
\end{definition}

Remember that the elliptical stadium is defined by two parameters: $a$ and $h$.
The two half-ellipses have minor axes equal to 1 and major axes equal to
$a>1$, and the straight parts have length $2h$.

\begin{theorem} \label{teo:coneB}
If $1<a<\sqrt{4-2\sqrt{2}}$ and $h>2a^2\sqrt{a^2-1}$, then
$\{C(\varphi,\theta)\}$
is eventually strictly invariant under $DR$.
\end{theorem}

outline of the proof: If $(\varphi_0,\theta_0)\in M_i,\:i=1, 2$, the trajectory
associated to $(\varphi_0,\theta_0)$ has at least two consecutive reflections
at
the same half-ellipse, being $\varphi_0$ one of them. So, the orbit remains
on the integral curve $F=F(\varphi_0,\theta_0)$ and the trajectory has,
while staying on the same half-ellipse, a conic caustic.
Taking $$M(i, {\cal E})=\{(\varphi_0,\theta_0)\:\hbox{such that the caustic is
an ellipse}\}$$ and $$M(i, {\cal H})=\{(\varphi_0,\theta_0)\:\hbox{such that
the caustic is a hyperbola}\}, $$ we have $B_i=(M(i, {\cal E}))\bigcup
(M(i, {\cal H}))\bigcup U_i$, $i=1, 2$.

Our aim is to prove that for almost every $(\varphi_0,\theta_0)\in B$, with
$(\varphi_n,\theta_n)=R^n(\varphi_0,\theta_0)$,
we will have $DR^n(C(\varphi_0,\theta_0))\subset C(\varphi_n,\theta_n)$
and, for almost every $(\varphi_0,\theta_0)\in B$, we can find a
$k=k(\varphi_0,\theta_0)$ such that
$DR^k(C(\varphi_0,\theta_0))\subset intC(\varphi_k,\theta_k). $

The cones $C(\varphi,\theta)$ are bounded by the vertical pencil and by either
the horizontal or the caustic pencil. If $a<\sqrt 2$, the vertical pencil is
deviated to the right under $DR$ (lemma~\ref{pro:vert}). Since
$det(DR_{(\varphi_0,\theta_0)})>0$, all we have to do is to compare the
slope of the image of $X_c(\varphi_0,\theta_0)$ or $X_h(\varphi_0,\theta_0)$,
under $DR$, with the slope of $X_c(R(\varphi_0,\theta_0))$ or
$X_h(R(\varphi_0,\theta_0))$, depending on where $R(\varphi_0,\theta_0)$ is.
We have to look for the point of tangency with a conical caustic or for the
intersection with a half-osculating circle, forward or backward, in order to
apply Donnay's Focusing Lemma. All the possible situations are analysed in 7
propositions in the appendix, and we conclude that if
$1<a<\sqrt{4-2\sqrt{2}}$ and $h>2a^2\sqrt{a^2-1}$, the forward focusing point
appears before the backward one.

\section{ Main Results}

Take $S:{\cal A}\longrightarrow{\cal A}$ the map associated to the billiard on
the elliptical stadium.

${\cal A}=[0,2\pi+4h]\times (0,\pi)={\cal B}_1\bigcup {\cal F}_1\bigcup {\cal
B}_2\bigcup{\cal F}_2$, where: ${\cal B}_1=[0,\pi] \times (0,\pi)$,
${\cal F}_1=[\pi+h,\pi+2h] \times (0,\pi)$,
${\cal B}_2=[\pi+2h, 2\pi+2h]\times (0,\pi)$
and ${\cal F}_2=[2\pi+2h, 2\pi+4h] \times (0,\pi)$, the ${\cal B}_i$
corresponding to the reflections at the elliptical part and the ${\cal F}_i$
corresponding to the reflections at the straight part.

Let $A={\cal A}\setminus (W\bigcap {\cal A})$, where $W$ is
the measure zero set defined in subsection~\ref{cones}.
$A=B_1\bigcup F_1\bigcup B_2\bigcup F_2$,
$B_i={\cal B}_i\setminus (W\bigcap {\cal B}_i),\:i=1, 2$,
$F_i={\cal F}_i\setminus (W\bigcap {\cal F}_i),\:i=1, 2. $

As in subsection~\ref{zonasB}, let
$U_i
=\{ (t_0,\theta_0)\in  B_i \:\hbox{such that}\:S(t_0,\theta_0) \,\hbox{and} \,
S^{-1}(t_0,\theta_0) \not\in B_i\}$ and
$M_i=\{ (t_0,\theta_0)\in B_i \: \hbox{such that}\:\hbox{either}\:
S(t_0,\theta_0) \in B_i \: \hbox{or} \:S^{-1}(t_0,\theta_0) \in B_i\}$.

\begin{corollary} \label{coro:coneA}
For each $(t,\theta)\in A$, define the measurable cone field $\{C(t,\theta)\}
\subset TA$ by
\begin{itemize}
\item if $(t,\theta)\in M_1\bigcup M_2$, $C(t,\theta)=\{rX_v+sX_c,\,
r. s\geq 0\}$.
\item
if $(t,\theta)\in U_1\bigcup U_2$, $C(t,\theta)=\{rX_v+sX_h,\, r. s\geq 0\}$.
\item if $(t,\theta)\in F_1\bigcup F_2$, $C(t,\theta)=DS(C(S^{-1}(t,\theta)))$.
\end{itemize}
If $1<a<\sqrt{4-2\sqrt{2}}$ and $h>2a^2\sqrt{a^2-1}$, then $\{C(t,\theta)\}$
is eventually strictly invariant under $DS$.
\end{corollary}

Applying Wojtkowski's theorem, it follows that for those values of $a$ and $h$,
the Lyapunov exponents are non-vanishing $\mu$-almost everywhere. In
(\cite{kn:liv},\S 14. A), Liverani and Wojtkowski proved that the strictly
invariance of the cone-field implies the non-contraction property of its
vectors. And also that the other conditions on the dynamical behaviour of the
singularity curves (the vertical lines on the phase space, given by $W_{0}$,
$W_{\pi}$, $W_{\pi + 2h}$ and $W_{2 \pi +2h}$) have been satisfied. All those
additional conditions allow to prove that the system is not only chaotic, ie,
has non-vanishing Lyapunov exponents, but also ergodic and (see \cite{kn:mark},
sec. 4) is a K-system.

So, we have:
\begin{theorem}
\label{teo:main}
If $a<\sqrt{4-2\sqrt 2}$, then $\forall h>2a^2\sqrt{a^2-1}$, the map
associated to the elliptical stadium has non-vanishing Lyapunov exponents
$\mu$-almost everywhere, is ergodic and has the K-property.
\end{theorem}

\section{ Final Remarks}

Actually, the numerical simulations suggest much more about the dynamical
behaviour, as we present bellow.
\bfig
 \vspace{5truecm}  \includegraphics{fig6.ps}
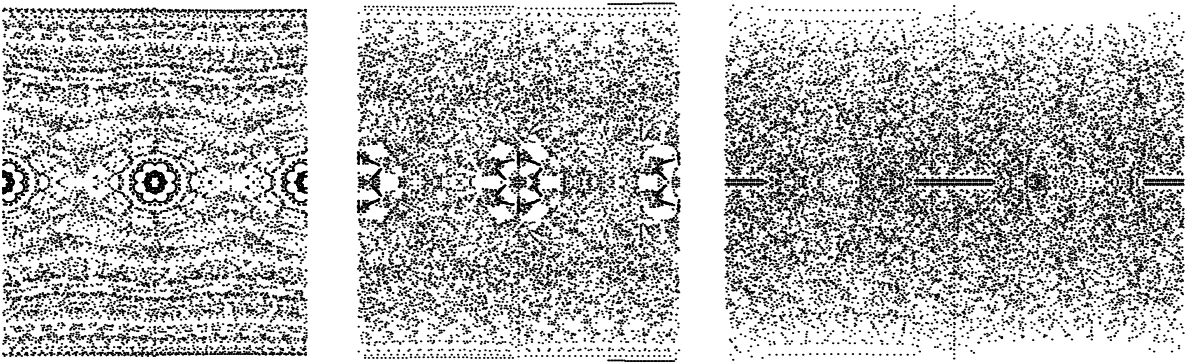
 \caption{$a=1.07$ and $h=0.01$, $h=0.1$, $h=0.815$}
\efig

1) $2a^2\sqrt{a^2-1}$ does not seem to be an optimal lower bound for $h$. The
phase space seems "chaotic" for values of $h$ smaller than this bound, at
least for $a<\sqrt{4-2\sqrt 2}$.

The dependence on $\sqrt{a^2-1}$ seems reasonable, since it measures the focal
distance of the half ellipses. What seems to be unnecessarily big is the
factor $2a^2$ that is clearly an overestimation for $h$, used on the proof
of proposition~\ref{pro:caso3}, to assure that the focusing points of
orbits from $M(1, {\cal H})$ to $M(2, {\cal H})$ are separated by a reflection
at the straight part.

In any case, the lower bound curve $h(a)=2a^2\sqrt{a^2-1}$ is in agreement with
what we expected for $a\sim 1$. For the circular stadium billiard, or $a=1$,
$h>0$ implies ergodicity. This will not be the case for $a\neq 1$, because the
structure of the phase space of the elliptical billiard is not as degenerate as
that of the circular billiard. The vertical tangent at $a=1$ reflects this
fact.

2) $\forall a>1$ and $\forall h>0$, it seems that a neighbourhood of the
boundaries $\theta=0$ and $\theta=\pi$ is invariant. As shown in
proposition~\ref{pro:caso2}, the behaviour of the cone field there is analogous
to
the behaviour of the cone field constructed by Donnay \cite{kn:don} for the
circular stadium. And this, perhaps, invariant neighbourhood is a chaotic zone
of positive measure, even for small $h$'s (figure~\ref{fig:uma orbita}).

3)  $\forall a>1$ and $\forall h>0$, some initial conditions near $\theta=0$
(or $\theta=\pi$), after a finite number of iterations, come near $\theta=\pi$
(resp. $\theta=0$), showing that the trajectory changes orientation in
relation to the orientation of the boundary of the billiard. This indicates
that the invariant manifolds of the hyperbolic 2-periodic orbit
$\{ (\pi/2,\pi/2), (3\pi/2+2h,\pi/2)\}$ (which have corners) cross and that
there are homoclinic points and, then, non-integrability
(figure~\ref{fig:uma orbita}).

\bfig
 \vspace{5truecm}  \includegraphics{fig7.ps}
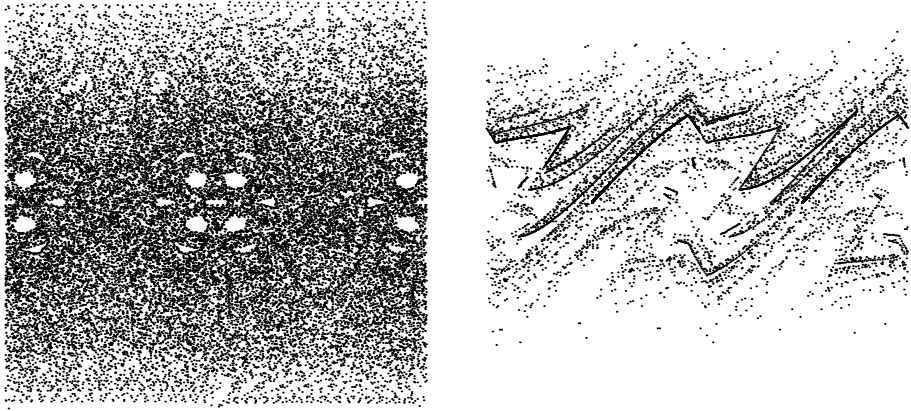
 \caption{32,000 points of a unique orbit and the unstable manifold,
        for  a=1. 07 and h=0.15}
 \label{fig:uma orbita}
\efig

4) Canale and Markarian \cite{kn:can} have proved the existence of symmetric
periodic orbits for the elliptical stadium and have pointed out that for
$a>1$, $h>0$, there exists a 4-periodic orbit, which is elliptic if
$h<\sqrt{a^2-1}$ and hyperbolic if $h>\sqrt{a^2-1}$.
The numerical simulations they caried out show that, while it is elliptic, it
is
encircled by elliptic islands of positive measure and, fixing $a<\sqrt 2$
and increasing $h$, those elliptic islands seem to be the last  to
disappear, among all those that can be observed, at least, for small values of
$h$. For $a>\sqrt 2$, the elliptic islands remain, even if $h$ is very big.

5) In the construction of the cone field, as in Donnay's proof \cite{kn:don},
it
becomes clear why the lower bound for $h$ must increase to infinity as $a$
approaches $\sqrt 2$. As $a$ approaches ${\sqrt 2}^{\: +}$, there is a
trajectory on $M\bigcap {\cal H}$ that has two reflections at the same
half-ellipse and cuts the minor axis of the half-ellipse closer and closer to
the center of the half-ellipse, ie, with a direction near the direction of
the asymptote of the hyperbolic caustic. This means that the focusing
distance of the caustic pencil goes to infinity and so, we need to put the
half-ellipses farther and farther apart to assure that the caustic pencil will
have enough space  to focus and to open again.

This also shows  that $h(a)=2a^2\sqrt{a^2-1}$ can not be the lower bound for
$a\sim \sqrt 2$. $h(a)$ must be determined by pieces. We believe
that the determination of $h(a)$ is closely related to the structure of
bifurcation of periodic orbits with hyperbolic caustic on the elliptical
billiard (figure~\ref{fig:bif}).

\appendix
\section*{Appendix}

We analyse, here, all the possibilities for a piece of orbit
$\{(\varphi_0,\theta_0), (\varphi_1,\theta_1)\}$.

\begin{proposition} \label{pro:caso1} 
Suppose that $(\varphi_1,\theta_1)=R(\varphi_0,\theta_0)
=T(\varphi_0,\theta_0)$, ie,
$\varphi_1$ and $\varphi_0$ are in the same half-ellipse. \\
If $1<a$ and $\forall h\geq 0$, then
$DR(C(\varphi_0,\theta_0))\subset C(\varphi_1,\theta_1)$, but the invariance is
not strict.
\end{proposition}
proof: If $(\varphi_1,\theta_1)=T(\varphi_0,\theta_0)$, then
$(\varphi_0,\theta_0)$ and $(\varphi_1,\theta_1)$ are on an integral curve
of the elliptical billiard and $(\varphi_0,\theta_0),\, (\varphi_1,\theta_1)
\in M_i,\: i=1, 2$. The cones are:
$$C(\varphi_j,\theta_j)=\{rX_v(\varphi_j,\theta_j)+sX_c(\varphi_j,\theta_j),\,
r. s\geq 0\},\: j=0, 1. $$
And
$DR_{(\varphi_0,\theta_0)}(X_c(\varphi_0,\theta_0))=DT_{(\varphi_0,\theta_0)}
(X_c(\varphi_0,\theta_0))=\lambda
X_c(\varphi_1,\theta_1),\:\lambda\neq 0$.

\begin{proposition} \label{pro:caso2} 
Suppose that \\ $(\varphi_0,\theta_0) \in M(1, {\cal E})$
(resp. $M(2, {\cal E})$) and $(\varphi_1,\theta_1)
\in M(2, {\cal E})$ (resp. $M(1, {\cal E})$). \\ \medskip
If $\forall a>1$ and $\forall h>0$, then
$DR(C(\varphi_0,\theta_0))\subset intC(\varphi_1,\theta_1)$.
\end{proposition}
proof: $(\varphi_0,\theta_0),\, (\varphi_1,\theta_1)
\in M_i,\, i=1, 2. $ and:
$$C(\varphi_j,\theta_j)=\{rX_v(\varphi_j,\theta_j)+sX_c(\varphi_j,\theta_j),\:
r. s\geq 0\},\: j=0, 1. $$
The caustic pencil $X_c(\varphi_0,\theta_0)$ focuses forward at $P_0$, the
point of tangency of the forward segment of the trajectory with the elliptical
caustic; and the caustic pencil $X_c(\varphi_1,\theta_1)$ focuses backward at
$P_1$, the point of tangency of the backward segment of the trajectory with
a new elliptical caustic. These two elliptical
caustics are no longer confocal, but have the same focusing distance
$c=\sqrt{a^2-1}$ and their foci lie on the same straight line.

If the trajectory between $\varphi_0$ and $\varphi_1$ does not hit the straight
part of the elliptical stadium, elementary geometry of ellipses
plus the focusing lemma give us that:
$$\hbox{slope}(DR_{(\varphi_0,\theta_0)}X_c(\varphi_0,\theta_0))>
\hbox{slope}(X_c(\varphi_1,\theta_1)). $$
If it hits the straight part, $P_0$ clearly appears before $P_1$ when
going from $\varphi_0$ to $\varphi_1$, and the focusing lemma gives us that:
$$\hbox{slope}(DR_{(\varphi_0,\theta_0)}X_c(\varphi_0,\theta_0))>
\hbox{slope}(X_c(\varphi_1,\theta_1)). $$

\begin{proposition} \label{pro:caso3} 
Suppose that either\\
a) $(\varphi_0,\theta_0) \in M(1, {\cal E})$ (resp. $M(2, {\cal E})$)
and $(\varphi_1,\theta_1) \in
M(2, {\cal H})$ (resp. $M(1, {\cal H})$),\\
or\\
b)  $(\varphi_0,\theta_0) \in M(1, {\cal H})$ (resp. $M(2, {\cal H})$)
and $(\varphi_1,\theta_1) \in
M(2, {\cal E})$ (resp. $M(1, {\cal E})$). \\
c)
$(\varphi_0,\theta_0) \in M(1, {\cal H})$
(resp. $M(2, {\cal H})$) and $(\varphi_1,\theta_1) \in
M(2, {\cal H})$ (resp. $M(1, {\cal H})$). \\
 \medskip
If $1<a<\sqrt{4-2\sqrt 2}$ and $h>2a^2\sqrt{a^2-1}$, then
$$DR_{(\varphi_0,\theta_0)}(C(\varphi_0,\theta_0))\subset
intC(\varphi_1,\theta_1) \ .$$
\end{proposition}
proof: of a) $(\varphi_1,\theta_1),\, (\varphi_0,\theta_0)
\in M_i,\, i=1, 2. $ And
$$C(\varphi_j,\theta_j)=\{rX_v(\varphi_j,\theta_j)+sX_c(\varphi_j,\theta_j),\,
r. s\geq 0\},\: j=0, 1. $$

The situation is equivalent to that of the last proposition, except that now
the caustic defined backward by $(\varphi_1,\theta_1)$ is a hyperbola.

The proof is based on the following results:

i) if $1<a<\sqrt 2$ and $(\varphi_1,\theta_1) \in M(i, {\cal H}),\:
i=1, 2$, then this part of the trajectory has exactly two reflections at the
same
half-ellipse, before leaving it (\cite{kn:don}).

ii) On the elliptical billiard, take a trajectory  on ${\cal H}$ having two
consecutive reflections $\varphi_0$ and $\varphi_1$ at the same half-ellipse.
If $1<a<\sqrt{4-2\sqrt 2}$, then the caustic pencils at $\varphi_0$ and
$\varphi_1$ focuses forward and backward inside the ellipse
(corollary~\ref{coro:DDD}).

iii) if $h>2a^2\sqrt{a^2-1}$, then every trajectory on $M_i\bigcap {\cal
H},\, i=1, 2$ has at least one reflection at the straight part of the
elliptical
stadium before and after the two consecutive reflections at the same
half-ellipse.
(proof: We should calculate the distance $d$ from $\varphi =0$ to the
reflection at the straight part of the elliptical stadium. The worst situation
is that of a trajectory that leaves $\varphi = 0$ in the direction of the
focus. Straightforward calculations lead to $d=4 c a^2$. )

The interior of the elliptical stadium may be seen as the union of the
interior of two ellipses, with half-axes $a$ and 1, and the interior of a
rectangle, with sides $2h$ and 2. Notice that, if $h>\sqrt{a^2-1}=c$, the foci
of the two ellipses are separated. Let us call $E_1$ and $E_2$ the interior of
the ellipses, and Q the interior of the rectangle. Notice that the straight
parts of the elliptical stadium are not contained in $E_1\bigcup E_2$.

{}From the results above, it is clear that the trajectory from $\varphi_0$ to
$\varphi_1$ touches an
elliptical caustic at $P_0$ in $E_1$ (or $E_2$), hits the straight part, and
then touches a hyperbolic caustic at $P_1$ in $E_2$ (or $E_1$).

The Focusing Lemma now gives that
$$\hbox{slope}(DR_{(\varphi_0,\theta_0)}X_c(\varphi_0,\theta_0))>
\hbox{slope}(X_c(\varphi_1,\theta_1)). $$

the proof of  b) is, by symmetry, equivalent to a);
the proof of c) is analogous.

\begin{proposition}\label{pro:caso6} 
Suppose that either\\
a) $(\varphi_0,\theta_0) \in M(1, {\cal E})$ (resp. $M(2, {\cal E})$)
and $(\varphi_1,\theta_1) \in U_2$ (resp. $U_1$)\\
or\\
b) $(\varphi_0,\theta_0) \in U_1$ (resp. $U_2$) and
$(\varphi_1,\theta_1) \in M(2, {\cal E})$ (resp. $M(1, {\cal E})$). \\
\medskip
If $1<a<\sqrt{2}$ and $h>\sqrt{a^2-1}$, then
$$DR_{(\varphi_0,\theta_0)}(C(\varphi_0,\theta_0))\subset
intC(\varphi_1,\theta_1). $$
\end{proposition}
proof: (of a)\, ) $(\varphi_0,\theta_0)\in M_i$, and $(\varphi_1,\theta_1)
\in U_i,\, i=1, 2. $ So, the cones are:
$$C(\varphi_0,\theta_0)=\{rX_v(\varphi_0,\theta_0)+sX_c(\varphi_0,\theta_0),\,
r. s\geq 0\}. $$
$$C(\varphi_1,\theta_1)=\{rX_v(\varphi_1,\theta_1)+sX_h(\varphi_1,\theta_1),\,
r. s\geq 0\}. $$

We have to compare the positions of the forward focusing point $P_0$ of
the pencil $X_c(\varphi_0,\theta_0)$ to the backward focusing point $P_1$ of
the pencil $X_h(\varphi_1,\theta_1)$. $P_0$ is the point of tangency with the
elliptical caustic and $P_1$ is the intersection of the backward segment of
trajectory with the half-osculating circle at $\varphi_1$.

If the trajectory between $\varphi_0$ and $\varphi_1$ hits the straight
part of the elliptical stadium, this reflection separates $P_0$ and $P_1$, and
the
Focusing Lemma gives
$$\hbox{slope}(DR_{(\varphi_0,\theta_0)}X_c(\varphi_0,\theta_0))>
\hbox{slope}(X_h(\varphi_1,\theta_1)). $$

Now suppose that it does not. Suppose also that $\varphi_0\in [\pi/2,\pi)$ (the
other cases being analogous by symmetry).

If $\varphi_0+\theta_0\geq\pi$, $P_0$ is contained in the interior of the
half-osculating circle of $\varphi=\pi$ (lemma~\ref{coro:tangelipse}).

On the other hand, call ${\cal O}_{\varphi_0}$ the union of the
half-osculating circles, from $\varphi=\varphi_0$ to
$\varphi=\varphi_0+\pi/2$. If $a<\sqrt 2$ and $h>c=\sqrt{a^2-1}$, ${\cal
O}_{\pi/2}$ and ${\cal O}_{3\pi/2}$ are disjoint and contained in the interior
of the elliptical stadium \footnote{
Given an ellipse with half--axis $1$ and $a$, $1<a<\sqrt 2$, centered at
$(0, 0)$, the intersection of ${\cal O}_{\pi/2}$ with the $x$-axis
$\subset [-\sqrt{a^2-1}, a]$ and of ${\cal O}_{3\pi/2}$ with the $x$-axis
$\subset [-a,\sqrt{a^2-1}]$. So, on the elliptical stadium, if
$h>\sqrt{a^2-1}$, ${\cal O}_{\pi/2}$ and ${\cal O}_{3\pi/2}$ are disjoint. }.
Then, if $\varphi_0\in [\pi/2,\pi),\,\varphi_0+\theta_0\geq \pi$ and
$\varphi_1\in [3\pi/2, 2\pi)$, the focusing lemma applies and
$$\hbox{slope}(DR_{(\varphi_0,\theta_0)}X_c(\varphi_0,\theta_0))>
\hbox{slope}(X_h(\varphi_1,\theta_1)). $$

If $\varphi_0\in [\pi/2,\pi),\,\varphi_0+\theta_0\geq \pi$ and
$\varphi_1\in (\pi, 3\pi/2)$, the segment of trajectory does not cross the
line containing the foci of the caustics. So, it touches a new elliptical
caustic on the other half-ellipse. Besides, $X_c(\varphi_1,\theta_1)$
has positive slope (because the integral curves on ${\cal E}$ are increasing
for $\varphi\in (\pi, 3\pi/2)$).

As it was shown in proposition~\ref{pro:caso2},
$$\hbox{slope}(DR_{(\varphi_0,\theta_0)}X_c(\varphi_0,\theta_0))>
\hbox{slope}(X_c(\varphi_1,\theta_1))$$
and $$\hbox{slope}(X_c(\varphi_1,\theta_1))>0=\hbox{slope}
(X_h(\varphi_1,\theta_1)). $$

If $\varphi_0+\theta_0<\pi$ and $h>c=\sqrt{a^2-1}$, $P_0$ is contained in the
interior of the half-ellipse (ie, $E_i\setminus (E_i\bigcap Q)$) and $P_1$ is
contained in ${\cal O}_{\pi}$.
So,
$\hbox{slope}(X_c(\varphi_1,\theta_1))>0=\hbox{slope}(X_h(\varphi_1,\theta_1))$.

The proof of b) is analogous to a) (by symmetry).

\begin{proposition} \label{pro:caso8} 
Suppose that either\\
a) $(\varphi_0,\theta_0) \in M(1, {\cal H})$ (resp. $M(2, {\cal H})$)
and $(\varphi_1,\theta_1) \in U_2$ (resp. $U_1$). \\
or\\
b) $(\varphi_0,\theta_0) \in U_1$ (resp. $U_2$) and
$(\varphi_1,\theta_1) \in M(2, {\cal H})$ (resp. $M(1, {\cal H})$). \\ \medskip
If $1<a<\sqrt{4-2\sqrt 2}$ and $h>2a^2\sqrt{a^2-1}$, then
$$DR_{(\varphi_0,\theta_0)}(C(\varphi_0,\theta_0))\subset
intC(\varphi_1,\theta_1). $$
\end{proposition}

proof: (of a) the proof is analogous to the first situation in
proposition~\ref{pro:caso6}, but notice that if
$(\varphi_0,\theta_0) \in M(1, {\cal H})$ (resp. $M(2, {\cal H})$)
and $(\varphi_1,\theta_1) \in U_2$ (resp. $U_1$), if
$1<a<\sqrt{4-2\sqrt 2}$ and $h>2a^2\sqrt{a^2-1}$, then the trajectory
touches a hyperbolic caustic at $P_0$ in $E_1$ (or $E_2$), hits the
straight part, and then cuts the half-osculating circle at $\varphi_1$.

The proof of b) is analogous to a) (by symmetry).

\begin{proposition} \label{caso10} 
Suppose that \\$(\varphi_0,\theta_0) \in U_1$ (resp. $U_2$) and
$(\varphi_1,\theta_1) \in U_2$ (resp. $U_1$). \\ \medskip
If $1<a<\sqrt 2$ and $h>\sqrt{a^2-1}$, then
$DR_{(\varphi_0,\theta_0)}(C(\varphi_0,\theta_0))\subset
intC(\varphi_1,\theta_1)$
\end{proposition}
proof: The cones are:
$$C(\varphi_i,\theta_i)=\{rX_v(\varphi_i,\theta_i)+sX_h(\varphi_i,\theta_i),\,
r. s\geq 0\}. $$
Suppose that $\varphi_0 \in [\pi/2,\pi)$ (the other cases being analogous by
symmetry). The forward focusing point $P_0$ of $X_h(\varphi_0,\theta_0)\in
{\cal O}_{\pi/2}$.

If $\varphi_1 \in [3\pi/2, 2\pi)$, the backward focusing point $P_1$ of
$X_h(\varphi_1,\theta_1)\in {\cal O}_{3\pi/2}$. Since $a<\sqrt 2$ and
$h>\sqrt{a^2-1}$, ${\cal O}_{\pi/2}$ and ${\cal O}_{3\pi/2}$ are disjoint,
the result follows.

If  $\varphi_1 \in (\pi, 3\pi/2]$, the trajectory does not cut the line
containing the foci. So, the trajectory touches backward an elliptic
caustic on the other half-ellipse at $\tilde P_1$. As in
proposition~\ref{pro:caso6},
$$\hbox{slope}(DR_{(\varphi_0,\theta_0)}X_h(\varphi_0,\theta_0))>
\hbox{slope}(X_c(\varphi_1,\theta_1)). $$
But $X_c(\varphi_1,\theta_1)$ has positive slope (because the integral curves
are increasing for $\varphi\in (\pi, 3\pi/2)$). So
$$\hbox{slope}(DR_{(\varphi_0,\theta_0)}X_h(\varphi_0,\theta_0))>
\hbox{slope}(X_c(\varphi_1,\theta_1))>0=\hbox{slope}
(X_h(\varphi_1,\theta_1)). $$

\noindent
{\bf Acknowledgements}. We thank profs. Eduardo Canale (Fac. Ing. ,
Montevideo), M\'ario Jorge Carneiro, Paulo C\'esar Carri\~ao, Susana Fornari
(ICEx/UFMG) and Alain Chenciner (Univ. Paris VII) for the good ideas they gave
us. We thank SBM (Sociedade Brasileira de Matem\'atica), FAPEMIG
(Funda\c c\~ao de Amparo \`a Pesquisa do Estado de Minas Gerais) and PrPq-UFMG
(Pr\'o-Reitoria de Pesquisa) for sponsoring the visit of RM to the Universidade
Federal de Minas Gerais.

\vfill
{
\small
\hfill
\begin{minipage}[t]{5truecm}
{\sc Roberto Markarian}\\
IMERL, Facultad de Ingenier\'\i a \\
C.C. 30 \\
Montevideo, Uruguay. \\
email: roma@imerl.edu.uy
\end{minipage}
\hfill
\begin{minipage}[t]{5truecm}
{\sc Sylvie Oliffson Kamphorst}\\
Departamento de Matem\'atica\\
ICEx--UFMG, C.P. 702\\
30161-970 Belo Horizonte, Brasil. \\
email: syok@mat.ufmg.br
\end{minipage}
\hskip 0.4truecm
\begin{minipage}[t]{5truecm}
{\sc S\^onia Pinto de Carvalho}\\
Departamento de Matem\'atica\\
ICEx--UFMG, C.P. 702\\
30161-970 Belo Horizonte, Brasil. \\
email: icedo@vm1.lcc.ufmg.br
\hskip 0.4truecm
\end{minipage}
\hfill
}

\end{document}